# Observation of Shubnikov de Haas and Aharanov-Bohm oscillations in silicon nanowires

Tahir Aslan, Davie Mtsuko, Christopher Coleman, Siphephile Ncube and Somnath Bhattacharyya [a]

Nano-Scale Transport Physics Laboratory, School of Physics, University of Witwatersrand, P/Bag 3, Johannesburg, 2050, South Africa

We record fine oscillations of 20 to 60 mT superimposed on larger oscillations having periodicity ~ 2 T at temperatures up to 100 K and fields up to 10 T from silicon nanowires. Having confirmed that these features appear from the edge states associated with skipping orbits at nanowire edges and confined pure orbits in the interior of the nanowires we derive electron effective mass of 0.001 $m_e$ to 0.006 $m_e$, carrier lifetime in the range 3 to 19 fs and carrier density that varies from $2\times10^{11}$ $cm^{-2}$ to $9\times10^{12}$ $cm^{-2}$. However, at low temperature the observed oscillation amplitude invariant of the field is attributed to not only a strong size confinement and the pinning of orbits by impurities but also Aharanov Bohm (AB) oscillations due to edge-states that propagate quasi-ballistically through the nanowire. The overall oscillation on a linear positive magnetoresistance background can be attributed to temperature-dependent crossover of Shubnikov de Haas oscillations (SdHO) and AB oscillations in silicon nanowires.

---

[a] Author to whom correspondence should be addressed. Electronic mail: Somnath.Bhattacharyya@wits.ac.za

# I. Introduction

Silicon is a standard material for fabrication of electronic devices. Silicon nanowires (SiNWs) exhibit interesting transport phenomena such as resonant tunneling, gate-dependent conductance modulation and sensitivity to adsorbed chemical species [1-3]. Owing to these transport properties SiNWs can find applications in high speed electronics, field effect transistor devices as well as chemical and biological sensors [4-6]. To make useful nano-electronic devices out of SiNWs we need to know various device parameters not only from DC measurements but also high magnetic field measurements. Silicon nanowires (SiNWs) are also a good material in which transport physics of quasi-one dimensional systems can be explored. This has driven lots of research works on electron transport properties of SiNWs [7-12]. The transport in SiNWs can be regarded as quasi-one dimensional depending on the diameter. It is believed that SiNWs with diameter greater than 20 nm should exhibit bulk-like transport due to many occupied subbands whereas in smaller diameter nanowires quantum confinement effects are significant [13]. Notable works in which quantum transport features have been observed include the work of Yi *et al.* on confinement induced quantum oscillations in zero-magnetic field room-temperature transport from sub-5 nm SiNWs that supported the theoretical predictions of enhanced mobility in strongly confined SiNW devices [9]. Zero-field quantum oscillation features have also been observed at low temperature in silicon quantum wire transistors [10, 11]. It will be interesting to study the magnetic field dependence of these SiNW systems in order to measure various transport parameters such as carrier concentration effective mass and carrier lifetimes that can help in establishing a complete picture of transport dynamics in SiNWs as well as provide transport parameters for designing of nanowire-based devices. A few works have been reported on magnetic field-dependent transport in SiNWs including the work in which anisotropic negative magnetoresistance (NMR) was observed in heavily phosphorous doped SiNWs [14], however no MR oscillations were observed. The NMR was attributed to electron scattering by localized spins at phosphorous donor states. Druzhinin *et al*. explored magnetoresistance in large diameter Si (5 µm), Ge and Si-Ge wires (20 µm) [15]. A large positive MR was observed in these nanowires but no oscillations were observed. However, we note that studies of magnetoresistance in small diameter SiNWs [14] have been rare although there are many works on nanowires of different materials such as Ge [16] and GaAs [17]. In this paper we present our studies of oscillatory magnetoresistance of undoped ~50 nm wide SiNWs. To the best of our knowledge, no significant claim on MR oscillations in silicon nanowires has been made so that one can derive useful transport parameters. There exist some works on Shubnikov de Haas oscillations (SdHO) and Aharanov-Bohm (AB) oscillations in nanowires made of other materials such as bismuth but not reported in SiNWs [18-20].

Another challenge for device is to record quantum effects including SdHO at elevated temperatures. However, owing to the large diameter of our silicon nanowires (~50 nm) such quantum effects may be modulated or smoothed out by superposition of effects in multiple subbands inherent in bulky nanowires. In such a case irregular oscillations may arise from the modulation of the SdHO by other kinds of oscillations known as magnetointersubband oscillations (MISO) and phonon-induced resistance oscillations whose origin can explained by scattering-assisted coupling of carrier states in different Landau levels [21,22]. In this paper we analyze the oscillatory MR in thick SiNWs and find the various device parameters such as electron effective mass, scattering time and concentration of defects and carriers.

## II. EXPERIMENTAL METHODS

The nanowires studied in this work were synthesized using laser ablation technique in which nickel was used as a growth catalyst. Purification and sonication was performed to remove excess catalyst particles and other irregular silicon microparticles. The purification was done with different acids and ultra-sonication in ethanol solvent for dispersion of nanowires. The four-probe devices were developed using polymethylmethacrylate (PMMA) lift-off based electron beam lithography. The electrodes were made of RF sputtered 4 nm titanium layer and 40 nm gold layer. The nanowires were characterized using scanning electron microscopy (SEM) (Fig.1 (a)), electrical transport measurements (Fig. 1 (b) and (c)), atomic force microscopy (AFM), Raman spectroscopy techniques (not shown here) and transmission electron microscopy (TEM) (Fig. 1 (d)).

In this study DC resistance measurements were performed in high vacuum in the temperature range 2.5 K to 300 K and magnetoresistance in the magnetic field range 0 to 10 T at different temperatures in a Cryogenics High Field measurement system interfaced with a Keithly source-meter and a nanovoltmeter. In the four-probe measurement a current of 1 to 10 μA is passed through the nanowire through terminals 1 and 4 and a voltage drop is read across the terminals 2 and 3. This takes out the contribution of resistance from the measurement leads as would be the case with a two-probe measurement

## III. RESULTS AND DISCUSSION

*Electrical characterization at zero magnetic field:*

Fig. 1 (b) shows the *I-V* characteristics of sample A (*D*~45 nm, *L*~3 µm) which are strongly non-linear with threshold voltages > 10 V on the positive bias side and < -10 V on the negative bias side. Kinks are also observed in the *I-V* curves at 2.7 K and 5 K. This is a signature of resonant tunneling conduction. It must also be pointed out that kinks in *I-V* characteristics have been theoretically predicted to occur in silicon nanowires which can be attributed to an interplay of quasiballistic transport and intersubband transitions of electrons due to elastic collisions with acoustic phonons, surface roughness and ionized impurities [23]. The current steps may also be related to an onset of negative differential resistance in the nanowire. Strong peaks are observed in the plot of differential conductance versus bias voltage (Fig. 1 (b) inset). Fig. 1 (c) shows the *I-V* characteristics of sample B (*D*~55 nm, *L*~1.5 µm) which are linear at room temperature and weakly non-linear at low temperature. The activation energy $E_A$ for sample A is estimated to be ~40 meV from low-temperature *R-T* measurements in which resistance scales with temperature as $R \propto exp(E_A/2k_BT)$ where $k_B$ is the Boltzmann constant. The small value of activation energy can be attributed to impurity states in the SiNW causing the reduction of gap which will be probed by the presence of magnetic field.

*Magnetoresistance study:*

We examine the magnetoresistance due to a magnetic field applied perpendicular to the SiNW. Magnetoresistance is defined as $MR = \frac{R(B) - R(0)}{R(0)}$ where *R(B)* is resistance which change with field and *R(0)* is the resistance at zero field. There are two notable features of the magnetoresistance shown in Fig. 2 (a) - (d). Firstly, a non-saturating linear positive MR is observed as shown in Fig. 2 (a) for a 3 µm long, 55 nm wide silicon nanowire. Here the resistance of the sample increases linearly as the magnetic field is swept from 0 to about 10 T. The slopes of MR lines are 0.005, 0.013 and 0.021 for 100 K, 10 K and 2.5 K, respectively. The increase in slope with the decrease of temperature indicates that charge carriers are more easily localised into cyclotron orbits at lower temperature. The positive MR is due to the creation of increasingly smaller cyclotron orbits as the magnetic field is increased, this has the effect of increasing the resistivity of the sample. As can be seen in Fig. 2 (a), the linear background is greatly reduced as the temperature is increased.

In Fig. 2 (b) for a 1.5 µm long, 45 nm wide nanowire a saturating positive MR is observed. The saturation of the MR with increase in field can be attributed to the limiting of the extent to which the cyclotron radius can be reduced. This is related to flux quantization that is to say there is a limit to the smallest flux that can enclosed by an orbiting electron.

The second noteworthy feature is that of oscillations in all the magnetoresistance plots. This oscillatory behaviour persists at temperatures 10 K and 100 K. The oscillations are irregular (with beating) more pronounced in sample A at 100 K with approximate period of ~ 2 T for major peaks and ~ 60 mT for the finer oscillations. These periods correspond to cyclotron radii of ~18 nm and ~104 nm. It is not hard to imagine the prevalence of circular 18 nm orbits within a 45 nm wide nanowire. However a circular 104 nm orbit cannot fit into a 45 nm nanowire unless it is of elliptical or other irregular shape. Therefore we may not associate the 60 mT oscillations to magnetic field induced circular cyclotron orbits in the bulk of the wire. The phase coherence length associated with the periodicity, estimated using formula $\Delta B = \varphi_0/L_\varphi^2$, is found to be 32 nm and 185 nm (for $\Delta B$~2 T and 0.06 T respectively) at 100 K [24]. Dividing the phase coherence lengths by the Fermi velocity ($1 \times 10^7$ m/s) we obtain quantum lifetimes in the range 3 to 19 fs. This is significantly small compared to scattering lifetime of 1 ps of bulk silicon [13] and indicates either the presence of large number of impurities or a large extent of surface roughness scattering in silicon nanowires.

From the magnetoresistance oscillations we can extract the effective mass of carriers in the silicon nanowires. To do this, we first plot the resistance as a function of $1/B$ as shown in Fig. 2 (c). The separation $\Delta(1/B)$ between SdH-like peaks is plotted as a function of $1/B$ minima in Fig 3 (a) inset. The carrier concentration $n$ is calculated from the formula $\Delta(1/B) = 2e/nh^2$ and plotted against $B$ in Fig 3 (a). The carrier concentration is found to scale linearly with $B$ for all temperature points. The carrier density values in the range $2\times10^{11}$ cm$^{-2}$ to $9\times10^{12}$ cm$^{-2}$ correspond to effective mass $m^* = (\frac{\hbar}{v_F})\sqrt{\pi n_s}$ that varies from 0.001 $m_e$ to 0.006 $m_e$ (where $m_e$ is the free electron mass) in the temperature range 2.5 K to 100 K and field range 1.5 T to 10 T. These $n_s$ values are comparable to carrier concentration of the order $10^{12}$ $cm^{-2}$ measured in multi-walled carbon nanotubes under transverse magnetic field [25]. The concentration of static scattering centers is estimated to be in the range 27 to 1200 cm$^{-2}$ using the slopes of the MR data at 2.5 K in Fig. 2 (a) and the generic description of linear quantum magnetoresistance given as

$$\rho_{xx} = \rho_{yy} = \frac{N_i H}{\pi n^2 e} \propto H, \qquad (1)$$

where $\rho_{xx}$ and $\rho_{yy}$ are the transverse components of the magnetoresistance, $N_i$ is the concentration of static scattering centers and $n$ is the density of electrons [26,27].

To extract the quantum lifetime we make use of the general form of magnetoresistance of a low dimensional system in the SdH oscillation regime which is given by [24]:

$$R_{XX} = R_0 \left[1 + \lambda \sum_{S=1}^{\infty} D(sX) exp\left(-\frac{s\pi}{w_c \tau}\right) \times \cos\left(s\frac{\hbar S_F}{eB} - s\pi + s\phi_0\right)\right]. \qquad (2)$$

From which to the first order we can write the amplitude of oscillations as: $A_{ex} = \lambda D(X) exp\left(-\frac{\pi}{w_c \tau}\right)$, where $w_c = eB/m^*$. The carrier lifetime can be calculated by plotting the amplitude of the oscillation versus the 1/$B$ at extrema of oscillations as shown in Fig. 3 (b) for 100 K. Using the slope of this graph and the value of effective mass obtained earlier we estimate the carrier life time as 13 fs. These values are of the same order of magnitude as quantum lifetimes measured in other low dimensional systems such as single layer graphene [28].

In the case of an applied perpendicular magnetic field it would be expected to observe harmonic SdH oscillations at low temperature as observed in graphene [28] and other 2DEG systems [29]. Usually in these oscillations the amplitude increases with increasing magnetic field due to smaller cyclotron orbits forming at higher fields as the Landau level crosses the Fermi level resulting in more of the edge state electrons being localized by the magnetic field. However, in our sample, the oscillation amplitude is independent of magnetic field and remains constant even at high fields. This anomalous behaviour can be explained as follows; Application of magnetic field perpendicular to the nanowire axis leads to formation of three different electron orbital motion: pure orbits, skipping orbits and also traversing orbits. In a 2D system and also in wide nanowires the pure orbits will become smaller and smaller as the magnetic field is increased thus leading to a conventional SdH oscillations with amplitude that increase with increasing field. However in narrow quantum wires the pure orbits in the core or bulk of the quantum wire are pinned by disorder or impurities. In other words, as the magnetic field is increased to a point that corresponds to a Landau level matching the Fermi level, edge states can be converted to cyclotron orbits. This conversion is however prevented due to backscattering resulting from impurities in the bulk of the nanowire. Due to this, an increase in the magnetic field does not necessarily lead to increasingly smaller cyclotron orbits to form, due to the electrons forming these orbits being scattered, and hence leads to oscillations of constant amplitude.

The lack of smearing or damping of the MR oscillation in Fig.2 a at temperature as high as 100 K prompts us to check whether the magnetoresistance oscillation may be just universal conductance fluctuations (UCF) which have been observed in a wide range of mesoscopic scale materials. UCFs are known to scale with length as $\Delta G = \frac{\Delta R}{\langle R \rangle} \approx \frac{e^2}{h}\left(\frac{l_\varphi}{L}\right)^{3/2}$ [30]. Using this relationship we obtain phase coherence lengths of 1.5 μm, 1 μm and 0.012 μm for 100 K, 10 K and 2.5 K, respectively. The decay of coherence with decreasing temperature is contrary to the expectation of improved coherence with lowering temperature in systems where phonons are the main phase breaking mechanism. Therefore these oscillations cannot be explained by UCF. It should be noted that high temperature magnetoresistance oscillations have before been observed in InGaAs quantum wells and it is not surprising to see the manifestation here [31]. These oscillations were attributed to transverse acoustic-phonon assisted transitions between magnetically induced states above the quantum well and a shallow impurity well.

We next consider the possibility of an AB effect origin of the oscillations for magnetic field oriented perpendicular to SiNWs. AB oscillations have been previously reported in multiwalled carbon nanotubes under magnetic field applied perpendicular to the nanotube but have not been reported in individual SiNWs [32]. Assuming that the large period oscillations (~2 T) observed in sample A are AB oscillations the area of the AB loops/rings for this period is $6.6\times10^{-16}$m$^2$ (i.e. a ring diameter of 52 nm). Such a ring diameter is comparable to the diameter of the nanowire determined by AFM (45 nm). A zoom into the conductance fluctuation of sample A (*D*~45 nm, *L*~3 um) [Fig. 2(d)] reveals fine Aharanov-Bohm-like oscillation of periods 20 mT. The 20 mT period correspond to an AB ring of area ($A=2\pi\hbar/e\Delta B$) of $2.1\times10^{-13}$ m$^2$ corresponding to a ring radius of 260 nm. AB loops of this area will only fit into a silicon nanowire if they were of elliptical shape with a minor axis of 45 nm and a major axis of at least 2.9 μm. A zoom into the conductance fluctuation of sample B (*D*~55 nm, *L*~1.5 μm) [Fig. 2(d)]) reveal fine AB-like oscillation of periods 60 mT. The AB oscillation may arise from interaction between the magnetically induced pure orbits in the core of the nanowire and propagating edge states at the surface of the nanowires [13]. The 60 mT period correspond to an AB ring of area of $6.9\times10^{-14}$ m$^2$ corresponding to a ring radius of 140 nm. AB loops of this area will only fit into a silicon nanowire if they were of elliptical shape with a minor axis of 55 nm and a major axis of at least 1.6 μm. The lengths of the major axes for AB ellipses (2.9 μm and 1.6 μm) are comparable to the lengths of the silicon nanowire device (3 μm and 1.5 μm). Thus, we may then associate the oscillation periods 20 mT and 60 mT to the AB phase of edge states which propagate ballistically through the wire and only scatter near the contacts. Using the relation $\Delta(1/B) = 2\pi e/\hbar A_{FS}$ we find that the range of $\Delta(1/B)$ 0.011 to 0.3 T$^{-1}$ corresponds to the cross sectional area of a Fermi surface pocket ($A_{FS}$) in the range $3.2\times10^{16}$ to $8.7\times10^{17}$ m$^{-2}$ in k-space. However the exact morphology of the Fermi surface pockets is not understood.

## IV. CONCLUSION

In this report we show a weakly linear positive magnetoresistance with oscillations which are predominantly AB-like at low temperature and SdH-like at high temperature in silicon nanowires. The calculated effective mass associated with oscillation extrema varies from 0.001 $m_e$ to 0.006 $m_e$, which corresponds to carrier concentration of $2\times10^{11}$ $cm^{-2}$ to $9\times10^{12}$ $cm^{-2}$ and the carrier lifetime in the range 3 to 19 fs, respectively. The intriguing existence of transverse MR

AB oscillation in silicon nanowires can be attributed to interference of edge-states that propagate ballistically in the nanowire. The weak temperature and magnetic field dependence of the oscillations can be attributed to quantum interference effects between skipping orbits, which penetrate into the core as an occupied Landau level is moved through the Fermi level, and pure cyclotron orbits pinned throughout the nanowire by impurity scatterers. These results show that the SiNWs synthesized in this work are of high mobility comparable to carbon nanotubes and graphene.

## ACKNOWLEDGEMENTS

SB acknowledges the CSIR-NLC for establishing the laser ablation set up which produced the silicon nanowires and NRF (SA) NNEP as well as nanotechnology flagship grant.

**Figure captions:**

**Fig. 1. (a)** SEM image of a four-terminal SiNW device ( sample B) developed using electron beam lithography. The diameter of the nanowire is 55 nm. The sample is contacted by a 4 nm titanium layer and a 40 nm gold layer. **(b)** I-V characteristics of sample A at various temperatures. Inset:  differential conductance as a function of bias voltage from I-V curve at 2.7 K that exhibited step-like and oscillatory features. **(c)**  DC current-voltage characteristics of the device shown in **(a)**.**(d)** TEM image of a typical silicon nanowire synthesized in this work.

**Fig 2 (a)** Transverse magnetoresistance of sample A (diameter 45 nm and length 3 μm) at various temperatures 2.5 K (black line), 10 K (red line) and 100 K (blue line). The MR curves have been offset by 0.5, 0.3 and 0.1 % for clarity. Oscillations are observed in all the magnetoresistance curves. **(b)** A saturating positive magnetoresistance is observed in sample B with diameter 55 nm and length 1.5 μm. **(c)** Variation of resistance with 1/B for sample A at 2.5 K, 10 K and 100 K. Note: The curves in **(c)** have been offset upwards for clarity. The amplitudes don't vary much with field and the oscillations are more AB like at 2.5 K. The oscillation amplitude are more pronounced in the intermediate and higher field regions at 10 K. The oscillations appear SdH-like with strong amplitude at higher field. **(d)** AB-like oscillations in the field range 2.5 to 4 T for sample A (black curve) and sample B (blue curve).

**Fig 3. (a)** The variation carrier density with magnetic field at various temperatures in sample A. Inset: Variation of peak spacing with 1/B minimum used to calculate the carrier density. **(b)** The variation of amplitude of extrema points of the resistance oscillations versus 1/B used to extract the quantum lifetime.

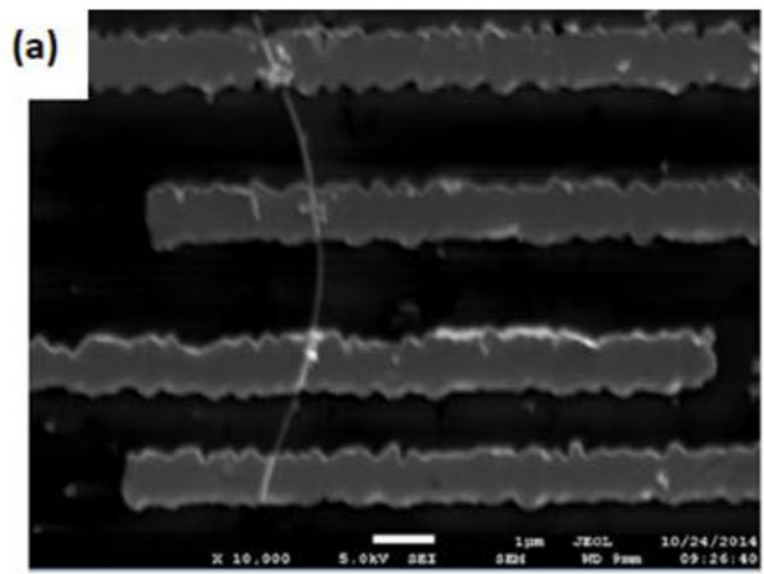
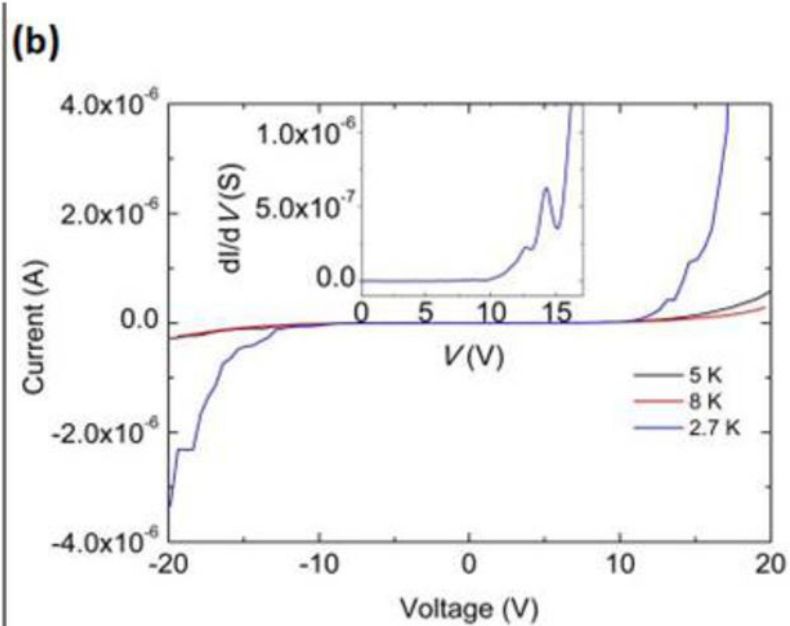
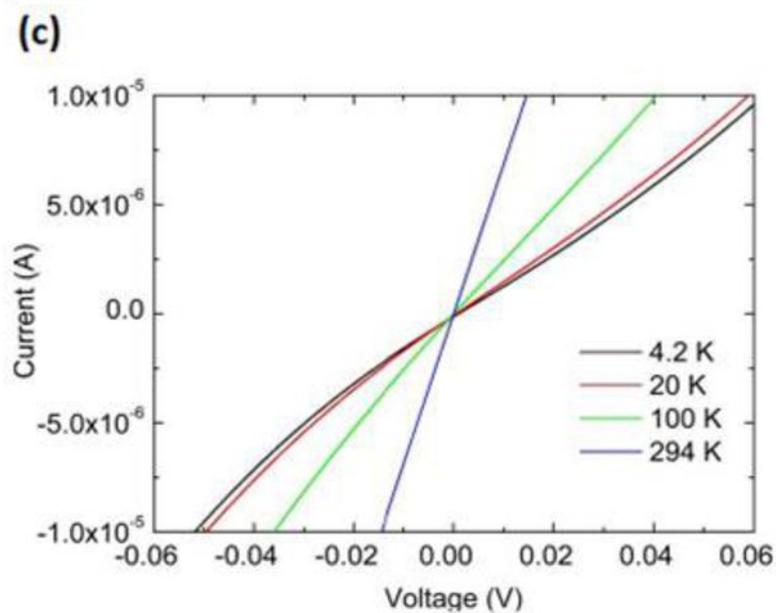
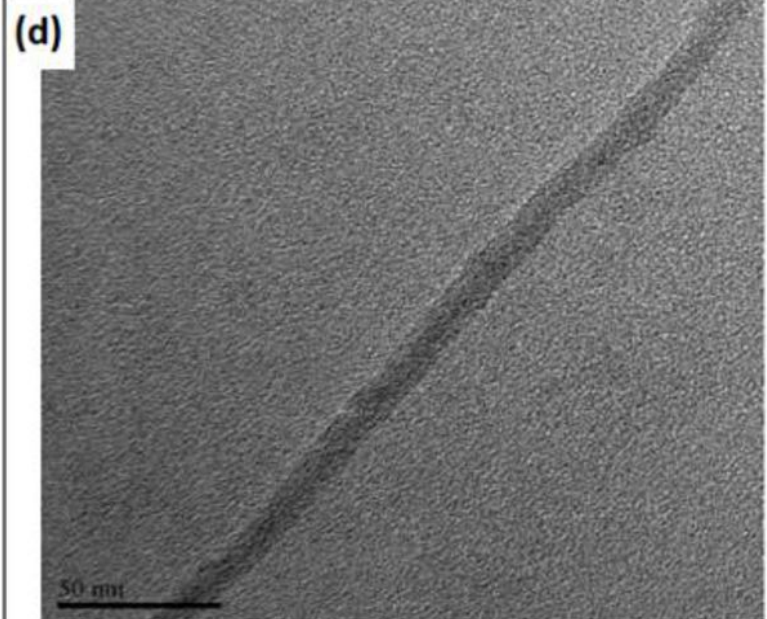

Fig 1. (a) - (d)

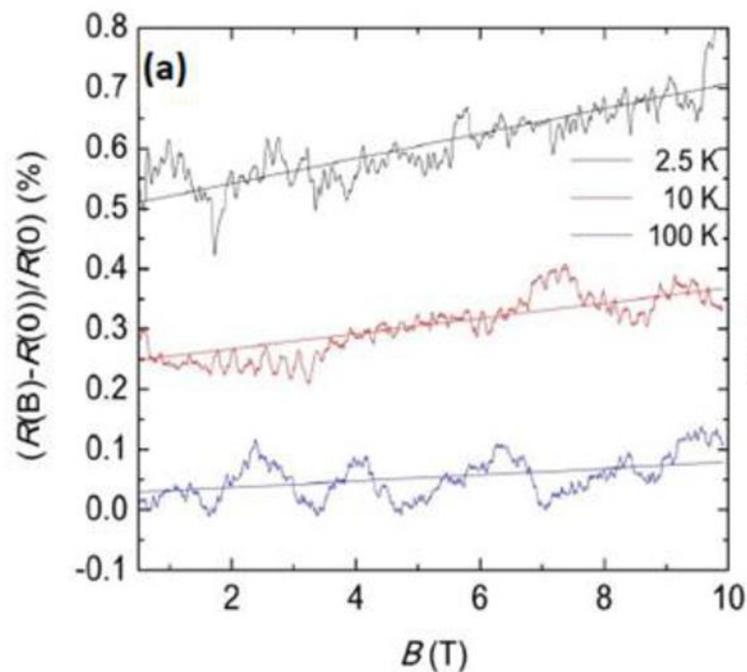
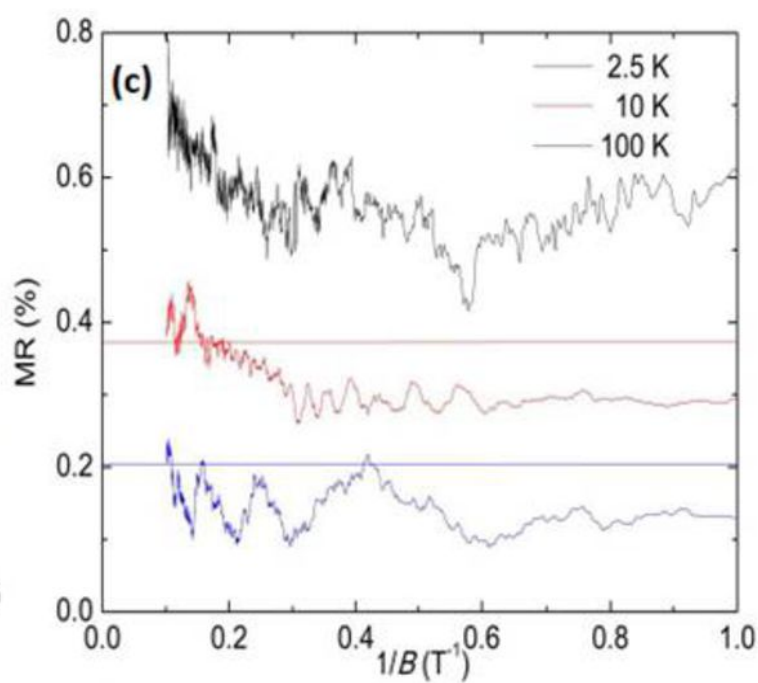
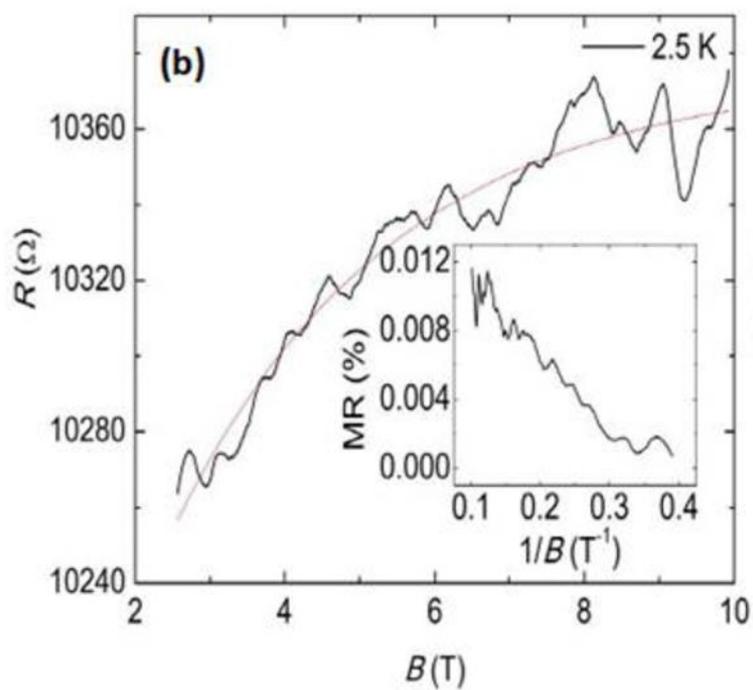
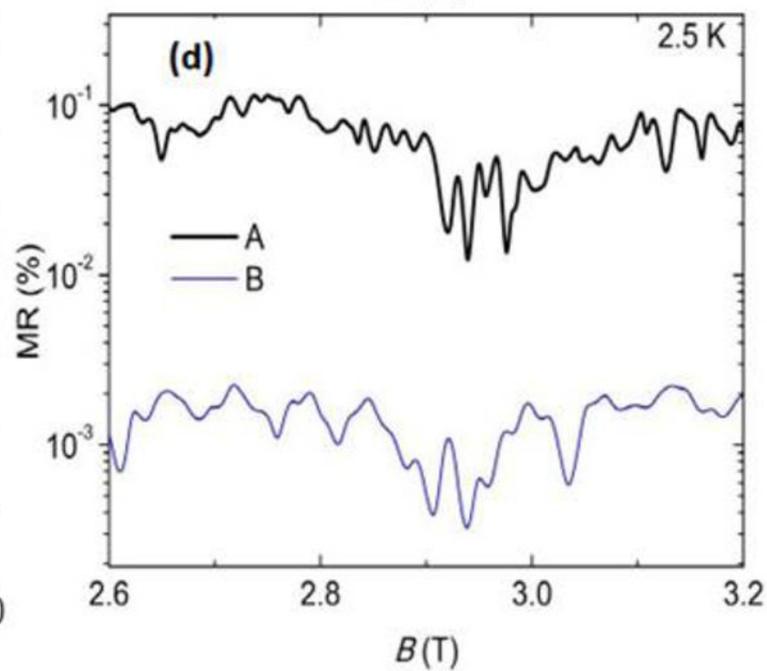

Fig 2. (a) - (d)

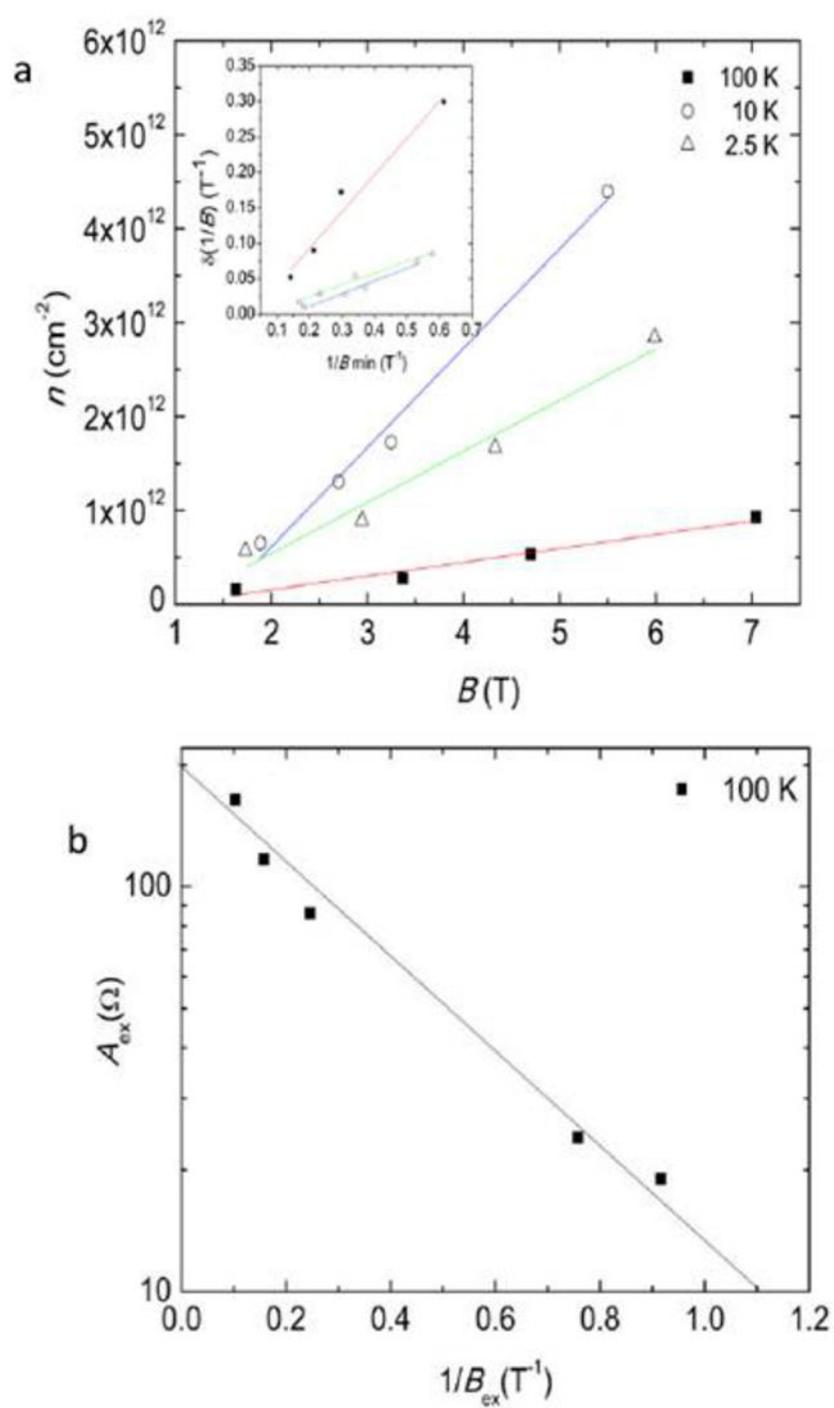

Fig 3. (a) and (b)